\documentclass[pra,twocolumn,showpacs]{revtex4}
\usepackage{amsmath,amssymb,mathrsfs}
\usepackage{psfrag}
\usepackage{graphicx}
\usepackage{graphics}
\usepackage{epsfig}
\usepackage{bm}
\usepackage{color}
\usepackage{verbatim,color,ulem}

\newcommand{\beq}{\begin{equation}}
\newcommand{\eeq}{\end{equation}}
\newcommand{\beqa}{\begin{eqnarray}}
\newcommand{\eeqa}{\end{eqnarray}}

\begin{document}

\title{Scattering length of composite bosons in the 3D BCS-BEC crossover}
\author{L. Salasnich$^{1,2,3}$ and G. Bighin$^{1,4}$}
\affiliation{$^{1}$Dipartimento di Fisica e Astronomia ``Galileo Galilei'', 
Universit\`a di Padova, Via Marzolo 8, 35131 Padova, Italy
\\
$^{2}$Consorzio Nazionale Interuniversitario per le Scienze Fisiche 
della Materia (CNISM), Unit\`a di Padova, Via Marzolo 8, 35131 Padova, Italy
\\
$^{3}$Istituto Nazionale di Ottica (INO) del Consiglio Nazionale 
delle Ricerche (CNR), Sezione di Sesto Fiorentino, Via Nello Carrara, 
1 - 50019 Sesto Fiorentino, Italy
\\
$^{4}$Istituto Nazionale di Fisica Nucleare (INFN), Sezione di Padova, 
Via Marzolo 8, 35131 Padova, Italy}

\date{\today}

\begin{abstract}
We study the zero-temperature grand potential of a 
three-dimensional superfluid made of ultracold fermionic 
alkali-metal atoms in the BCS-BEC crossover. In particular, we analyze 
the zero-point energy of both fermionic single-particle excitations 
and bosonic collective excitations. The bosonic elementary excitations, 
which are crucial to obtain a reliable equation of state 
in the BEC regime, are obtained with a 
low-momentum expansion up to the forth order of the quadratic (Gaussian) 
action of the fluctuating pairing field. By performing a 
cutoff regularization and renormalization of Gaussian fluctuations, 
we find that the scattering length $a_B$ 
of composite bosons, bound states of fermionic pairs, 
is given by $a_B=(2/3) \ a_F$, where $a_F$ is 
the scattering length of fermions. 
\end{abstract}


\maketitle

\section{Introduction}

Some years ago the crossover from the weakly-paired 
Bardeen-Cooper-Schrieffer (BCS) state to the Bose-Einstein 
condensate (BEC) of molecular dimers 
was experimentally achieved using ultracold fermionic alkali-metal atoms 
in a three-dimensional (3D) configuration 
\cite{exp_MolecularBEC,exp_Crossover,chin}. 
These remarkable experimental results have been   
investigated by several theoretical groups 
(for a recent review see \cite{book-bcs}). 
In a recent experiment the BCS-BEC crossover 
has been analyzed also in a quasi-2D atomic gas \cite{exp3}, 
where the experimental data are not in full agreement with theoretical 
predictions \cite{randeria2d,bertaina}. In a very 
recent paper \cite{sala-composite} we have 
studied Gaussian fluctuations of the zero-temperature 
attractive Fermi gas in the 2D BCS-BEC crossover 
showing that they are crucial to get a reliable equation of state 
in the BEC regime of composite bosons, bound states of fermionic pairs. 
In particular, performing dimensional regularization and 
renormalization-group analysis, we have found the remarkable analytical 
formula $a_B = a_F/(2^{1/2}e^{1/4})$ which connects 
the scattering length $a_B$ of composite bosons 
to the scattering length $a_F$ of fermions \cite{sala-composite}. 

In this paper we investigate the relationship between $a_B$ and 
$a_F$ in the 3D BCS-BEC crossover. We study the low-momentum expansion, 
up to the fourth order, of the quadratic action of the fluctuating 
pairing field finding that it gives an ultraviolet divergent 
contribution of the Gaussian fluctuations to the grand potential. 
We perform a cutoff regularization of this zero-point energy 
in the BEC regime and obtain the simple but meaningful analytical 
formula $a_B = (2/3)\ a_F$. 
This result is in good agreement with other beyond-mean-field 
theoretical predictions: $a_B\simeq 0.75\ a_F$ 
based on a diagrammatic approach \cite{pieri}, 
$a_B \simeq 0.60\ a_F$ derived from a four-body analysis \cite{petrov} 
and also from Monte Carlo simulations \cite{astra}, 
and $a_B\simeq 0.55 \ a_F$ obtained with convergence 
factors \cite{hu,randeria2}. We stress, however, that 
our result is fully analytical, contrary to all other 
beyond-mean-field predictions \cite{pieri,petrov,astra,hu,randeria2}, 
and it is based on a transparent cutoff regularization and subsequent 
renormalization of bare physical parameters. 

\section{The model}

We consider a three-dimensional Fermi gas of ultracold and dilute 
two-spin-component neutral atoms. The atomic fermions 
are described in the path integral formalism by the complex Grassmann 
fields $\psi_{s} ({\bf r},\tau )$, $ \bar{\psi}_{s} ({\bf r},\tau )$ 
with spin $s = ( \uparrow , \downarrow )$  \cite{nagaosa,atland}. 
The partition function ${\cal Z}$ of the uniform 
system at temperature $T$, in a three-dimensional box of volume $V$, 
and with chemical potential $\mu$ can be written as 
\beq 
{\cal Z} = \int {\cal D}[\psi_{s},\bar{\psi}_{s}] 
\ \exp{\left\{ -{1\over \hbar} \ S  \right\} } \; , 
\eeq
where 
\beq 
S = \int_0^{\hbar\beta} 
d\tau \int_{V} d^3{\bf r} \ \mathscr{L}
\eeq
is the Euclidean action functional and  $\mathscr{L}$ is 
the Euclidean Lagrangian density, given by 
\beq 
\mathscr{L} = \bar{\psi}_{s} \left[ \hbar \partial_{\tau} 
- \frac{\hbar^2}{2m}\nabla^2 - \mu \right] \psi_{s} 
+ g \, \bar{\psi}_{\uparrow} \, \bar{\psi}_{\downarrow} 
\, \psi_{\downarrow} \, \psi_{\uparrow} 
\eeq
where $g$ is the strength of the s-wave inter-atomic 
coupling ($g<0$ in the BCS regime) \cite{nagaosa,atland}. 
Summation over the repeated index $s$ in the Lagrangian is meant and  
$\beta \equiv 1/(k_B T)$ with $k_B$  Boltzmann's constant. 

The Lagrangian density $\mathscr{L}$, which is 
quartic in the fermionic fields, 
can be rewritten as a quadratic form  by introducing the
auxiliary complex scalar field $\Delta({\bf r},\tau)$ so that 
\cite{nagaosa,atland}:
\beq 
{\cal Z} = \int {\cal D}[\psi_{s},\bar{\psi}_{s}]\, 
{\cal D}[\Delta,\bar{\Delta}] \ 
\exp{\left\{ - {S_e(\psi_s, \bar{\psi_s},
\Delta,\bar{\Delta}) \over \hbar} \right\}} \; , 
\eeq
where 
\beq 
S_e(\psi_s, \bar{\psi_s},\Delta,\bar{\Delta}) = \int_0^{\hbar\beta} 
d\tau \int_{{V}} d^3{\bf r} \ 
\mathscr{L}_e(\psi_s, \bar{\psi_s},\Delta,\bar{\Delta})
\label{e-action}
\eeq
and the effective Euclidean Lagrangian 
density $\mathscr{L}_e(\psi_s, \bar{\psi_s},\Delta,\bar{\Delta})$ reads 
\beq 
\mathscr{L}_e =
\bar{\psi}_{s} \left[  \hbar \partial_{\tau} 
- {\hbar^2\over 2m}\nabla^2 - \mu \right] \psi_{s} 
+ \bar{\Delta} \, \psi_{\downarrow} \, \psi_{\uparrow} 
+ \Delta \bar{\psi}_{\uparrow} \, \bar{\psi}_{\downarrow} 
- {|\Delta|^2\over g} \; . 
\label{ltilde}
\eeq 
In this paper we investigate the effect of 
fluctuations of the gap field $\Delta({\bf r},t)$ around its
mean-field value $\Delta_0$ which may be taken to be real. 
For this reason we set 
\beq 
\Delta({\bf r},\tau) = \Delta_0 +\eta({\bf r},\tau)  \; , 
\label{polar}
\eeq
where $\eta({\bf r},\tau)$ is the complex paring field of bosonic 
fuctuations \cite{nagaosa,atland,stoof}. 

\section{Mean-field}

By setting $\eta({\bf r},t)=0$, 
and integrating over the fermionic fields  $\psi_s({\bf r},t)$ 
and $\bar{\psi}_s({\bf r},t)$ 
one obtains the mean-field (saddle-point and 
fermionic single-particle) partition function \cite{nagaosa,atland,stoof}
\beq 
{\cal Z}_{mf} =  \exp{\left\{ - {S_{mf}\over \hbar} \right\}}
= \exp{\left\{ - \beta \, \Omega_{mf} \right\}} \; , 
\eeq
where 
\beqa 
{S_{mf}\over \hbar} &=& - Tr[\ln{(G_0^{-1})}] - 
\beta {V} {\Delta_0^2\over g} \; 
\nonumber
\\
&=& - \sum_{{\bf k}} \left[ 2
\ln{\left( 2 \cosh{(\beta E_{sp}(k)/2)} \right)} 
- \beta (\epsilon_k -\mu) \right] 
\nonumber
\\
&-& \beta {V} {\Delta_0^2\over g} \; , 
\label{omega-sp} 
\eeqa
with $\epsilon_k=\hbar^2k^2/(2m)$, 
\beq 
G_0^{-1} = \left(
\begin{array}{cc}
\hbar \partial_{\tau} -{\hbar^2\over 2m}\nabla^2 -\mu & \Delta_0 \\ 
\Delta_0 & \hbar \partial_{\tau} +{\hbar^2\over 2m}\nabla^2 +\mu
\end{array}
\right)
\label{G0}
\eeq
the inverse mean-field Green function, and 
\beq 
E_{sp}(k)=\sqrt{(\epsilon_k-\mu)^2+\Delta_0^2} 
\label{ex-fermionic}
\eeq 
the energy of the fermionic single-particle elementary excitations. 

At zero temperature ($T=0$, i.e. $\beta\to +\infty$) 
the mean-field grand potential $\Omega_{mf}$ becomes 
\beq 
\Omega_{mf} = -\sum_{\bf k} \left( E_{sp}(k) - \epsilon_k + \mu \right) 
- V {\Delta_0^2\over g} \; . 
\label{omega0-div}
\eeq
The constant, uniform and real gap parameter $\Delta_0$ is obtained 
by minimizing $\Omega_{mf}$ with respect to $\Delta_0$, namely
\beq 
\left({\partial \Omega_{mf}\over \partial \Delta_0}\right)_{\mu} = 0 \; . 
\eeq 
In this way one gets the familiar gap equation 
\beq
-{1\over g} = {1\over {V}} \sum_{|{\bf k}|<\Lambda} {1\over 2E_{sp}(k)} \; ,  
\label{gap-div}
\eeq 
where the ultraviolet cutoff $\Lambda$ is introduced to avoid the 
divergence of the right side of Eq. (\ref{gap-div}) 
in the continuum limit $\sum_{\bf k}\to V\int d^3{\bf k}/(2\pi)^3$. 
One can express the bare interaction strength $g$ in terms of 
the physical s-wave scattering length $a_F$ 
of fermions through \cite{randeria,marini,schakel}
\beq 
- {m\over 4\pi\hbar^2 a_F} = - {1\over g} - 
{1\over {V}} \sum_{|{\bf k}|<\Lambda} {1\over 2 \epsilon_k } \; . 
\label{magic}
\eeq
After integation over momenta Eq. (\ref{magic}) reads
\beq 
- {m\over 4\pi\hbar^2 a_F} = - {1\over g} - 
{m\over 2\pi^2\hbar^2} \Lambda \; . 
\label{magic2}
\eeq
For this expression one deduces that while $g$ is always negative 
$a_F$ can change sign. In particular, 
in the weak-coupling BCS limit, where $g\to 0^-$, the first term 
on the right of Eq. (\ref{magic2}) dominates 
and $a_F=mg/(4\pi\hbar^2)\to 0^-$. 
In the strong-coupling BEC limit, where $g\to -\infty$, the second 
term on the right of Eq. (\ref{magic2}) dominates and $a_F=\pi/(2\Lambda) 
\to 0^+$ when $\Lambda$ is sent to infinity \cite{schakel}. 

Inserting Eq. (\ref{magic}) into Eq. (\ref{gap-div}) we obtain 
the regularized gap equation 
\beq
-{m\over 4\pi\hbar^2a_F} = {1\over {V}} \sum_{|{\bf k}|<\Lambda} 
\left( {1\over 2E_{sp}(k)} - {1\over 2\epsilon_k} \right) \; ,  
\label{gap-r}
\eeq 
where one can safely take the limit $\Lambda\to+\infty$, 
finding the energy gap $\Delta_0$ as a function 
of the chemical potential $\mu$ and the scattering length $a_F$. 
We stress that in the BCS limit, where $a_F\to 0^-$, 
the chemical potential 
$\mu$ is positive and $\mu/\Delta_0 \to +\infty$. At unitarity, where 
$a_F\to \pm \infty$, Eq. (\ref{gap-r}) gives 
$\mu/\Delta_0=0.8604$ showing that the chemical potential $\mu$ is 
still positive. In the BEC regime, where $a_F\to 0^+$, 
the chemical potential becomes negative and it is given by 
\beq 
\mu = -{\hbar^2\over 2ma_F^2} + {1\over 4} 
{ma_F^2\over \hbar^2} \Delta_0^2 \; , 
\label{echem-bec}
\eeq
while $\mu/\Delta_0 \to -\infty$. Notice that $\epsilon_B=\hbar^2/(m a_F^2)$ 
is the binding energy of the atomic dimers (composite bosons) 
which are formed at unitarity \cite{randeria,marini,schakel} 
and clearly $\mu = -\epsilon_B/2$ in the deep BEC limit. 

By using Eq. (\ref{magic}) (with $\Lambda\to +\infty$) 
with Eq. (\ref{omega0-div}) we obtain 
the zero-temperature 
regularized mean-field grand potential of fermionic single-particle 
excitations 
\beq 
\Omega_{mf} = 
-\sum_{\bf k} \left( E_{sp}(k) - \epsilon_k +\mu 
- {\Delta_0^2\over 2\epsilon_k} 
\right) - V {m\over 4\pi\hbar^2 a_F}{\Delta_0^2} \; . 
\label{omega0-r}
\eeq
In the BCS limit, where $a_F\to 0^-$, the energy gap $\Delta_0$ goes 
to zero and the mean-field grand potential becomes that of a non-interacting 
Fermi gas, namely 
\beq
\Omega_{mf} = - V \ {2 \over 15 \pi^2}
\left({2m\over \hbar^2}\right)^{3/2} \mu^{5/2} \; .
\label{omega0-bcs}
\eeq
In the BEC limit, where $a_F\to 0^+$, the mean-field grand potential 
reads \cite{randeria}
\beq 
\Omega_{mf} = - V \, {1 \over 256 \pi} 
\left({2m\over \hbar^2}\right)^{3/2} {\Delta_0^4\over |\mu|^{3/2}} \; .  
\label{omega0-bec}
\eeq  
We point out that in this limit both $\Delta_0$ and $|\mu|$ ($\mu <0$) diverge 
but $|\mu|$ diverges faster and the grand potential $\Omega_{mf}$ 
goes to zero. 

\section{Gaussian quantum fluctuations}

We now consider the effect of quantum fluctuations, i.e. in Eq. (\ref{polar})
we allow $\eta({\bf r},t)\neq 0$. 
Expanding the effective action $S_e(\psi_s, \bar{\psi_s},\Delta,\bar{\Delta})$ 
of Eq. (\ref{e-action}) around $\Delta_0$ 
up to the quadratic (Gaussian) order in $\eta({\bf r},t)$ 
and $\bar{\eta}({\bf r},t)$ one finds 
\beq 
Z = Z_{mf} \ \int 
{\cal D}[\eta,\bar{\eta}] \ 
\exp{\left\{ - {S_g(\eta,\bar{\eta}) \over \hbar} \right\}} \; , 
\label{sigo}
\eeq
where 
\beq 
S_{g}(\eta,\bar{\eta}) = {1\over 2} \sum_{q} 
({\bar\eta}(q),\eta(-q)) \ {\bf M}(q) \left(
\begin{array}{c}
\eta(q) \\ 
{\bar\eta}(-q) 
\end{array}
\right) \; 
\eeq 
is the Gaussian action of fluctuations in the reciprocal space 
with $q=({\bf q},i\nu_m)$ 
the $4$-vector denoting the momenta ${\bf q}$ and Bose Matsubara 
frequencies $\nu_m=2\pi m/\beta$. The $2\times 2$ matrix ${\bf M}(q)$ 
is the inverse fluctuation propagator, whose non trivial dependence 
on $q$ can be found in Ref. \cite{ohashi2,randeria2}. The energy 
$E_{col}(q)$ of the bosonic collective excitations 
can be extracted from $M(q)$ 
\cite{ohashi2,randeria2,schakel,tempere2,marini2} and it is given by 
\beq 
E_{col}(q) = 
\sqrt{\epsilon_q \left( \lambda \ \epsilon_q + 2 \ m \ c_s^2 \right)} 
\eeq
where $\epsilon_q=\hbar^2q^2/(2m)$ is the free-particle energy, 
$\lambda$ takes into account the first correction 
to the familiar low-momentum phonon dispersion 
$E_{col}(q) \simeq c_s \hbar q$, $c_s$ being the sound velocity. 
Both $\lambda$ and $c_s$ depend on the chemical potential $\mu$ and 
the energy gap $\Delta_0$. In particular, one finds \cite{marini,marini2}
\beq 
\lambda = 
\left\{ 
\begin{array}{cc} 
-{128\over 135} {\mu^2\over \Delta_0^2} & \mbox{BCS limit}
\\ 
{1\over 8} & \mbox{unitarity}
\\
{1\over 4} & \mbox{BEC limit}
\end{array}
\right. 
\label{lambda}
\eeq
and 
\beq 
m \ c_s^2 = 
\left\{ 
\begin{array}{cc} 
{2\over 3} \mu & \mbox{BCS limit}
\\ 
{2\over 3} \mu & \mbox{unitarity}
\\
{1\over 8}{\Delta_0^2\over |\mu|} & \mbox{BEC limit}
\end{array}
\right. 
\label{cs}
\eeq

Integrating over the bosonic fields $\eta(q)$ 
and $\bar{\eta}(q)$ in Eq. (\ref{sigo}), at zero temperature we find 
the grand potential 
\beq 
\Omega = - \lim_{\beta \to +\infty} {1\over \beta} \ln{(Z)} 
= \Omega_{mf} + \Omega_{g} \; 
\eeq
where $\Omega_{mf}$ is given by Eq. (\ref{omega0-r}), 
while $\Omega_{g}$ reads
\beq 
\Omega_{g} = {1\over 2} \sum_{|{\bf q}|<\Lambda} E_{col}(q) \; . 
\label{omegacol-div}
\eeq
This is the zero-point energy of bosonic collective 
excitations \cite{randeria,ohashi2,schakel}. 
Also here the ultraviolet cutoff $\Lambda$ is introduced to avoid the 
divergence in the continuum 
limit $\sum_{\bf q}\to V\int d^3{\bf q}/(2\pi)^3$. 

\section{Scattering length of composite bosons in the BEC limit}

Expanding Eq. (\ref{omegacol-div}) in powers of $\Lambda$ \cite{schakel} 
we find  
\beqa 
{\Omega_{g}\over V} &=& 
{\hbar^2\over 40\pi^2 m \lambda^{1/4}} \Lambda^5 
+ {m c_s^2 \over 12\pi^2 \lambda^{1/2}}\Lambda^3  
\nonumber
\\
&-& {m^3c_s^4\over 4\pi^2 \lambda^{3/2}} \Lambda + 
{8 m^4c_s^5\over 15\pi^2 \hbar^2 \lambda^{2}} + O({1\over \Lambda}) . 
\label{piopio}
\eeqa

In this equation the first two terms are truly divergent;
the third term, despite being $\propto \Lambda$ is
indeed convergent, as we are going to show shortly and
the fourth term is subleading in the BEC limit.

We then want to regularize the $\Lambda^5$ and $\Lambda^3$
terms and this can be done by redefining the bare parameters
in the mean-field grand potential in eq. (\ref{omega0-bec}),
introducing their renormalized counterparts. One then finds,
as shown in detail in Appendix \ref{app}, that the following 
regularization pattern removes the divergencies at the leading order:
\beqa
\begin{cases}
\Delta_0 \longrightarrow \Delta_0 + \frac{16 \sqrt{2}}{5} 
\frac{|\mu|^\frac{3}{2}}{\Delta_0^3} 
\left( \frac{\hbar^2}{2m} \right)^{\frac{5}{2}} \Lambda^5 \\
|\mu| \longrightarrow |\mu| - \frac{128}{9 \pi^3} 
\frac{|\mu|^\frac{1}{2}}{\Delta_0^2} 
\left( \frac{\hbar^2}{2m} \right)^\frac{5}{2} \Lambda^5
\end{cases}
\label{pattern}
\eeqa having used the deep-BEC relations in 
Eqs. (\ref{lambda}), (\ref{cs}) and, as 
previously discussed, the fact that in the strong-coupling BEC limit 
the cutoff $\Lambda$ can be obtained 
from Eq. (\ref{magic2}) and it reads
\beq
\Lambda = {\pi\over 2 a_F} \; .
\label{supermagic}
\eeq

We now analyze the leading convergent contribution, i.e. the $\Lambda$ 
term of Eq. (\ref{piopio}); even though it seems divergent, being proportional
to $\Lambda$, actually it is not because in the BEC limit 
$c_s^4$ goes to zero faster than $1/\Lambda$. 

Using Eqs. (\ref{echem-bec}), (\ref{lambda}), (\ref{cs}),  
(\ref{supermagic}) in the deep BEC regime we have 
$|\mu|=\hbar^2/(2ma_F^2)$, $\lambda=1/4$, $mc_s^2=\Delta_0^2/(8|\mu|)$, 
and the $\Lambda$ term of Eq. (\ref{piopio}) becomes 
\beq 
\Omega_{g} = - V \ {\alpha \over 256 \pi} 
\left({2m\over \hbar^2}\right)^{3/2} {\Delta_0^4\over |\mu|^{3/2}} \; ,  
\label{omegacol-bec}
\eeq
with $\alpha =2$. It is important to stress that Eq. (\ref{omegacol-bec}) 
is formally the same formula found by Hu, Liu, and Drummond \cite{hu} 
and also by Diener, Sensarma, 
and Randeria \cite{randeria2} by using a different regularization 
procedure based on convergence factors. They have determined 
numerically the parameter $\alpha$ finding 
respectively $\alpha=2.5$ \cite{hu} and$\alpha=2.61$ \cite{randeria2}, 
while here we derive $\alpha=2$ analytically. 

In the BEC limit, taking into account Eqs. (\ref{omega0-bec}) 
and (\ref{omegacol-bec}) the total grand potential is given by 
\beq 
\Omega = \Omega_{mf} + \Omega_{g} = 
- V {(1+\alpha) \over 256 \pi} 
\left({2m\over \hbar^2}\right)^{3/2} {\Delta_0^4\over |\mu|^{3/2}} \; ,  
\label{omega-bec}
\eeq
with $(1+\alpha)=(1+2)=3$. Here the grand potential $\Omega$ 
depends explicitly on both $\mu$ and 
$\Delta_0$. Consequently, the total number $N$ of fermions must be 
calculated as follows 
\beq 
N = - \left( 
{\partial \Omega\over \partial \mu}\right)_{V,\Delta_0} - 
\left({\partial \Omega\over \partial \Delta_0}\right)_{V,\mu} 
{\partial \Delta_0\over \partial \mu} \; ,  
\label{stanco}
\eeq 
and the number density $n=N/V$ reads 
\beq 
n = {(1+\alpha)\over 16 \pi} \left({2m\over \hbar^2}\right)^{3/2} 
{\Delta_0^2 \over |\mu|^{1/2}}  \; . 
\label{n-bec}
\eeq 
Notice that in Eq. (\ref{stanco}) the second term accounts for a variation 
of the order parameter $\Delta_0$ as a function of $\mu$ 
instead of considering it as an independent variable. 
The same procedure was adopted in Refs. \cite{hu,randeria2}, while 
in the well-known Nozieres-Schmitt-Rink scheme this term is absent. 
As stressed in Refs. \cite{hu,randeria2}, this second term 
cannot be neglected as 
it becomes more and more relevant in the BEC regime we consider here, 
becoming of the same order of the first term. 

Eq. (\ref{n-bec}) shows that, at fixed number density $n$, 
in the BEC limit, where both $|\mu|$ and $\Delta_0$ go to infinity, 
one has $|\mu| \sim \Delta_0^4$ and, from Eq. (\ref{omega-bec}) it 
follows that $\Omega$ goes to zero. 
To obtain Eq. (\ref{n-bec}) we have used Eq. (\ref{stanco}) but also 
Eq. (\ref{echem-bec}), 
which immediately gives ${\partial \Delta_0/\partial \mu} 
= {2\hbar^2/(m a_F^2 \Delta_0)}\simeq 4|\mu|/\Delta_0$. Taking into 
account Eq. (\ref{n-bec}), the equation (\ref{echem-bec}) 
for the chemical potential in the BEC limit can be rewritten as 
\beq 
\mu = -{\hbar^2\over 2ma_F^2} + {\pi \hbar^2\over m} 
{a_F\over (1+\alpha)} n \; , 
\eeq
where the second term is half of the chemical 
potential $\mu_B=4\pi \hbar^2 a_B n_B/m_B$ 
of composite bosons of mass $m_B=2m$, density $n_B=n/2$, 
and boson-boson scattering length 
\beq 
a_B = {2\over (1+\alpha)} a_F = {2\over 3} a_F \; . 
\label{our}
\eeq
This result is in good agreement with other beyond-mean-field 
theoretical predictions: $a_B \simeq 0.75a_F$ of Pieri 
and Strinati \cite{pieri}, $a_B \simeq 0.60 a_F$ of Petrov, Salomon and 
Shlyapnikov \cite{petrov} (and also 
Astrakharchik, Boronat, Casulleras, and S. Giorgini \cite{astra}), 
and $a_B \simeq 0.55 a_F$ of Hu, Liu and Drummond \cite{hu} 
(and also Diener, Sensarma and Randeria \cite{randeria2}). 
On the other hand the mean-field result is quite different, 
namely $a_B=2a_F$ \cite{stoof}. As stressed in the introduction, 
the beauty of our result, Eq. (\ref{our}), is that it is obtained 
fully analytically, contrary to all other beyond-mean-field predictions 
\cite{pieri,petrov,hu,randeria2}. 

\section{Conclusions}

Starting from a theory of attractive fermions, performing cutoff 
regularization plus renormalization of Gaussian fluctuations, 
we have obtained a remarkable formula 
between the scattering length $a_B$ of composite bosons and the 
scattering length $a_F$ of fermions. This formula, $a_B=(2/3)\, a_F$, 
is based on the two-body scattering theory which gives in the 
deep BEC regime an explicit relationship 
between the cutoff parameter and the s-wave scattering length. 
Our approach, limited to the 
quartic term in the low-momentum expansion of bosonic collective 
excitations, is fully reliable in the BEC regime but it 
cannot describe the entire 3D BCS-BEC crossover. 
In fact in the BCS region, where the chemical potential $\mu$ is positive, 
the sign of the coefficient $\lambda$ of the collective spectrum 
is negative (pair instability) and 
further terms must be included in the momentum expansion. 

\section*{Acknowledgments}

The authors acknowledge for partial support 
Ministero Istruzione Universita Ricerca (PRIN project 
2010LLKJBX) and thank Pieralberto Marchetti, 
Adriaan Schakel and Flavio Toigo for useful suggestions. 

\appendix*
\section{Renormalization in the BEC limit}
\label{app}

By using Eqs. (\ref{lambda}), (\ref{cs}) and (\ref{supermagic}) we
rewrite the divergent terms in the Gaussian grand potential in
Eq. (\ref{piopio}) as a function of $\Delta_0$ and $\mu$

\beq 
\Omega_g^{(\Lambda^3)} = V \frac{\pi}{384} \left( \frac{2m}{\hbar^2} 
\right)^\frac{3}{2} \Delta_0^2 | \mu |^\frac{1}{2} 
\eeq \beq 
\Omega_g^{(\Lambda^5)} = V \frac{\sqrt{2} \pi^3}{640} 
\left( \frac{2m}{\hbar^2} \right)^\frac{3}{2} | \mu |^\frac{5}{2} 
\eeq
The tree-level mean-field contribution in Eq. (\ref{omega0-bec}) is
a function of the bare parameters $\Delta_0$ and $|\mu|$,
which we now renormalize as follows:

\beq
\begin{cases}
\Delta_0 \longrightarrow \Delta_R = \Delta_0 + \delta \Delta \\
|\mu| \longrightarrow |\mu|_R = |\mu| + \delta \mu
\end{cases}
\eeq for a suitable choice of the counter-terms. The renormalized
mean-field grand potential now reads, to the leading order:

\begin{widetext}
\begin{equation}
\Omega_{mf}^R = - V \, {1 \over 256 \pi} 
\left({2m\over \hbar^2}\right)^{3/2} \frac{\Delta_R^4}{|\mu|_R^\frac{3}{2}} 
\approx - V \, {1 \over 256 \pi} 
\left({2m\over \hbar^2}\right)^{3/2} \left( \frac{\Delta_0^4}
{|\mu|^\frac{3}{2}} + 4 \frac{\Delta_0^3}{|\mu|^\frac{3}{2}} 
\delta \Delta - \frac{3}{2} \frac{\Delta_0^4}{|\mu|^\frac{5}{2}} 
\delta \mu \right)
\label{rgp}
\end{equation}
\end{widetext}
and in order to find the renormalization pattern in Eq. (\ref{pattern})
we simply equate the new terms in the renormalized grand potential in
Eq. (\ref{rgp}) to minus each divergent term, i.e.:

\beq
\frac{3}{2} V \, {1 \over 256 \pi} 
\left({2m\over \hbar^2}\right)^{3/2} \frac{\Delta_0^4}
{|\mu|^\frac{5}{2}} \delta \mu = - \Omega_g^{(\Lambda^3)}
\eeq
\beq
- 4 V \, {1 \over 256 \pi} 
\left({2m\over \hbar^2}\right)^{3/2} \frac{\Delta_0^3}
{|\mu|^\frac{3}{2}} \delta \Delta = - \Omega_g^{(\Lambda^5)}
\eeq
solving for the counter-terms one obtains $\delta \mu = 
-\frac{4 \pi ^2 |\mu|^3}{9 \Delta_0^2}$
and $\delta \Delta = \frac{\pi ^4 |\mu|^4}{5 \sqrt{2} \Delta_0^3}$.
Eq. (\ref{pattern}) readily follows reinstating
$\Lambda$ by means of Eqs. (\ref{lambda}), (\ref{cs}) and (\ref{supermagic}).

\end{document}